
\tolerance=10000
\input phyzzx


\REF\tim{ C.M. Hull, {\it Timelike T-duality, de Sitter space, large N gauge
theories and topological field theory},  JHEP {\bf 9807} (1998) 021,
hep-th/9806146.}
\REF\mn{J. Maldacena
and C. Nunez, {\it  Supergravity description of field theories on curved manifolds
and a  no go theorem} Int.J.Mod.Phys. {\bf A16} (2001) 822; hep-th/0007018.}
\REF\hms{ S.~Hawking, J.~Maldacena and A.~Strominger, {\it DeSitter entropy,
quantum entanglement and AdS/CFT},  JHEP {\bf 0105}, 001 (2001);
hep-th/0002145.} \REF\bousso{ R.~Bousso, {\it Bekenstein bounds in de Sitter
and flat space}, JHEP {\bf 0104}, 035 (2001); hep-th/0012052.}
\REF\jmas{ J.~Maldacena and A.~Strominger, {\it Statistical entropy of de
Sitter space}, JHEP {\bf 9802}, 014 (1998); gr-qc/9801096.}
\REF\banksb{ T.~Banks, {\it Cosmological breaking of supersymmetry or little
Lambda goes back to  the future. II}, hep-th/0007146.}
\REF\boussoc{ R.~Bousso, {\it Holography in general space-times}, JHEP {\bf
9906}, 028 (1999) [hep-th/9906022].}
\REF\vijay{ V.~Balasubramanian, P.~Horava and D.~Minic, {\it Deconstructing de
Sitter}, JHEP {\bf 0105}, 043 (2001) [hep-th/0103171].}
\REF\boussob{ R.~Bousso, {\it Positive vacuum energy and the N-bound}, JHEP
{\bf 0011}, 038 (2000) [hep-th/0010252].}
\REF\witten{ E.~Witten, {\it Quantum gravity in de Sitter space},
hep-th/0106109. }
\REF\strom{A. Strominger, hep-th/0106113.}
\REF\Odi{S.~Nojiri and S.~D.~Odintsov,
Phys.\ Lett.\ B {\bf 519}, 145 (2001), hep-th/0106191 and hep-th/0107134.}
\REF\klemm{D. Klemm, hep-th/0106247.}
\REF\gib{ G.W. Gibbons, \lq Aspects of Supergravity Theories', Published in {\it
GIFT Seminar on  Supersymmetry, Supergravity and Related Topics}, edited by
F. del Aguila, J. A. de Azcarraga and L. E. Ibanez, World Scientific (1984).
}
\REF\nct{C.M. Hull, {\it Phys. Rev.} {\bf D30} (1984) 760; C.M. Hull, {\it Phys.
Lett.} {\bf 142B} (1984) 39; C.M. Hull, {\it Phys. Lett.} {\bf 148B} (1984) 297;
C.M. Hull, {\it Physica} {\bf 15D} (1985) 230; Nucl. Phys. {\bf B253} (1985) 650; 
C.M. Hull, {\it Class. Quant. Grav.} {\bf 2} (1985) 343;
C.~M.~Hull and N.~P.~Warner,
Nucl.\ Phys.\ B {\bf 253}, 650 (1985)
and 
Nucl.\ Phys.\ B {\bf 253}, 675 (1985).}
\REF\gunwar{ M. G\"unaydin, L.J. Romans and N.P. Warner, {\it Gauged $N=8$
Supergravity in Five Dimensions}, Phys. Lett. {\bf 154B}, n. 4 (1985) 268; {\it
Compact and Non--Compact Gauged Supergravity Theories in Five Dimensions}, Nucl.
Phys. {\bf B272} (1986) 598.}
\REF\randsum{L. Randall and R. Sundrum, Phys. Rev. Lett. {\bf 83} (1999) 3370;
4690.}
\REF\gel{M. Gell-Mann and B. Zwiebach,  Phys. Lett. {\bf 141B} (1984) 333, {\bf
147B} (1984) 111; C. Wetterich, Nucl. Phys. {\bf 242} (1984) 473.}
\REF\CW{C.M. Hull and N. P. Warner, Class. Quant. Grav. {\bf 5} (1988) 1517.}
\REF\NicWet{
H.~Nicolai and C.~Wetterich,
{\it On The Spectrum Of Kaluza-Klein Theories With Noncompact Internal
Spaces}, Phys.\ Lett.\ B {\bf 150}, 347 (1985).}
\REF\nfor{A. Das, M. Fischler and M. Rocek, Phys. Rev. {\bf D16} (1977)
3427.}
\REF\adsstab{P. Breitenlohner and D.Z. Freedman, Phys. Lett. {\bf 115B} (1982)
197;  Ann. Phys. {\bf 144} (1982) 197; G.W. Gibbons, C.M. Hull and N.P. Warner,
Nucl. Phys. {\bf B218} (1983) 173.}
\REF\mal{J. Maldacena,  
Adv.Theor.Math.Phys. {\bf 2} (1998) 231-252; Int.J.Theor.Phys. {\bf 38} (1999)
1113-1133; hep-th/9711200.}
\REF\gat{S.J. Gates and B. Zwiebach,
Nucl.\ Phys.\ B {\bf 238}, 99 (1984);
Phys.\ Lett.\ B {\bf 123}, 200 (1983).}
\REF\vans{K. Pilch, P. van Nieuwenhuizen and M. Sohnius, Commun. Math. Phys. {\bf
98} (1985) 105.}
\REF\time{C. M. Hull, {\bf JHEP} 9811:017, 1998, hep-th/9807127.}
\REF\luk{J. Lukierski and A. Nowicki,   Phys.Lett. {\bf 151B}  (1985)  382.}
\REF\hlw{P.S. Howe, N.D. Lambert and P.C. West,  {\it A new massive Type IIA
supergravity from compactification}, Phys. Lett. {\bf B416} 303-308, (1998);
hep-th/9707139. }
\REF\pope{ I.V. Lavrinenko, H. Lu and C.N. Pope, {\it Fibre bundles and
generalised dimensional reduction}, Class.Quant.Grav. 15, 2239-2256 (1998);
hep-th/9710243.}
\REF\chamlam{A. Chamblin and N.D. Lambert, hep-th/0102159.}
\REF\CJ{E. Cremmer and B. Julia, Phys. Lett. {\bf 80B} (1978) 48; Nucl. Phys.
{\bf B159} (1979) 141.}
\REF\dwn{ B. de Wit and H. Nicolai,  {\it Phys. Lett.} {\bf 108 B} (1982) 285; B.
de Wit and H. Nicolai,  {\it Nucl. Phys.} {\bf B208} (1982) 323.}
\REF\frefo{ F. Cordaro, P. Fr\'e, L. Gualtieri, P. Termonia and M. Trigiante,
{\it $N=8$ gaugings revisited: an exhaustive classification}, Nucl. Phys. {\bf
B532} (1998) 245.}
\REF\PPV{ M. Pernici, K. Pilch and P. van Nieuwenhuizen, {\it Gauged
$N=8~D=5$ Supergravity} Nucl. Phys. {\bf B259} (1985) 460.}
\REF\popec{
M.~Cvetic, H.~Lu, C.~N.~Pope, A.~Sadrzadeh and T.~A.~Tran,
Nucl.\ Phys.\ B {\bf 590}, 233 (2000)
hep-th/0005137.}
\REF\HK{C. M. Hull and R. R. Khuri, Nucl. Phys. {\bf B536} (1999) 219,
hep-th/9808069.}
\REF\HKt{C. M. Hull and R. R. Khuri,  Nucl.Phys. {\bf  B575} (2000) 231-254,
hep-th/9911082.}
\REF\witt{E. Witten,  Adv.Theor.Math.Phys. {\bf 2} (1998) 253; hep-th/9802150.}
\REF\Bob{B.S. Acharya, M. O'Loughlin and B. Spence, Nucl.
Phys. {\bf B503} (1997) 657; B.S. Acharya, J.M. Figueroa-O'Farrill, M.
O'Loughlin and B. Spence,
hep-th/9707118.}
\REF\thom{M. Blau and G. Thompson, Phys. Lett. {\bf B415} (1997) 242.}
\REF\peet{
A.~W.~Peet and J.~Polchinski,
Phys.\ Rev.\ D {\bf 59}, 065011 (1999)
hep-th/9809022; H.~J.~Boonstra, K.~Skenderis and P.~K.~Townsend,
JHEP {\bf 9901}, 003 (1999), 
hep-th/9807137.}
\REF\dilres{ G.W. Gibbons, G.T. Horowitz and P.K. Townsend, Class. Quan.
Grav. {\bf 12} (1995) 297, hep-th/9410073.}
\REF\abz{M. Abou Zeid, C.M. Hull,  Nucl.Phys. {\bf B561} (1999) 293;
  hep-th/9907046.}


\font\mybb=msbm10 at 12pt
\def\bbbb#1{\hbox{\mybb#1}}

\def\R {\bbbb{R}}
 %

\def \aa {\alpha}
\def \bb {\beta}
\def \gg {\gamma}
\def \dd {\delta}

\def \ll {\lambda}
\def \mm {\mu}

\def \rr {\rho}
\def \ss {\sigma}
\def \tt {\tau}

\def \ddd {\Delta}

\def \lll {\Lambda}

\def \www{\Omega}

\def\sym {super Yang-Mills}

\def \ti {\tilde}

\def \2 {{1 \over 2}}
\def \3 {{1 \over 3}}
\def \4 {{1 \over 4}}
\def \5 {{1 \over 5}}
\def \6 {{1 \over 6}}
\def \7 {{1 \over 7}}
\def \8 {{1 \over 8}}
\def \9 {{1 \over 9}}
\def \0 { \infty}

\def\++ {{(+)}}
\def \- {{(-)}}
\def\+-{{(\pm)}}

\def \pa {\partial}

\def \qq {\qquad}


 \def\unit{\hbox to 3.3pt{\hskip1.3pt \vrule height 7pt width .4pt \hskip.7pt
\vrule height 7.85pt width .4pt \kern-2.4pt
\hrulefill \kern-3pt
\raise 4pt\hbox{\char'40}}}
\def\II{{\unit}}

\def\H {{\cal{H}}}

\def\nup#1({Nucl.\ Phys.\  {\bf B#1}\ (}



\Pubnum{ \vbox{  \hbox {QMUL-PH-01-09}  
\hbox{hep-th/0109213} 
}}
\pubtype{}
\date{September 2001}

\titlepage

\title {\bf  De Sitter Space in Supergravity and M Theory}

\author{C.M. Hull}
\address{Physics Department,
\break
Queen Mary, University of London,
\break
Mile End Road, London E1 4NS, U.K.}
\vskip 0.5cm

\abstract { Two   ways in which de Sitter space can arise in supergravity
theories are discussed. In the first, it   arises as a solution  of a
conventional supergravity, in which case it   necessarily has no Killing
spinors. For example, de Sitter space can arise  as a solution  of $N=8$ gauged
supergravities in four or five dimensions. These   lift to solutions
of 11-dimensional supergravity or $D=10$ IIB supergravity which are warped
products of de Sitter space and non-compact spaces of negative curvature. In
the second way, de Sitter space    can arise as a supersymmetric solution of
an unconventional supergravity theory, which typically has some kinetic terms
with the \lq wrong' sign; such solutions are invariant under a de Sitter
supergroup. Such solutions lift to supersymmetric solutions of unconventional
supergravities in $D=10$ or $D=11$, which nonetheless arise as field theory
limits of theories that can be obtained from M-theory by timelike T-dualities 
and related dualities. Brane solutions interpolate between these solutions and
flat space and lead to a holographic duality between theories in  de Sitter
vacua and Euclidean conformal field theories. Previous results are reviewed
and generalised, and discussion  is included of Kaluza-Klein theory with
non-compact internal spaces, brane and cosmological solutions, and holography
on de Sitter spaces and product spaces.  }

\endpage


\chapter{Introduction}

There has recently been renewed interest in quantum gravity in de Sitter space
and the issue of obtaining de Sitter space from string theory [\tim-\klemm]. Here
the ways in which de Sitter space can arise in supergravity theories will be
examined, and the relevance of these for string theory will be discussed.

In [\gib,\mn], it was argued that de Sitter space cannot arise from a conventional 
compactification of  supergravity, string or M-theory. If $d$-dimensional de Sitter
space is to emerge from string theory, it must therefore arise in some non-standard
way. However it arises, the desired result would be to have a low-energy description
in terms of a $d$-dimensional effective field theory in de Sitter space, and this
theory should have local supersymmetries, which  will typically all be
spontaneously broken by the de Sitter solution. The constraints of local
supersymmetry are  very restrictive, but nonetheless there are theories with 
maximal local supersymmetry in $d=4$ and $d=5$ dimensions with  de Sitter
solutions of this type [\nct,\gunwar]. These have a number of puzzling features,
but   the fact that there are so few examples consistent with maximal local
supersymmetry suggests these are worth closer examination. 

Given such a $d$
dimensional theory, the next question is whether it can be obtained from string
theory or M-theory, or from a solution of some
$D$-dimensional supergravity ($D>d$). In most of the known cases, the $D$
dimensional origin of these de Sitter theories is  as a solution  which is a
(possibly warped) product of $d$-dimensional de Sitter space and a $D-d$
dimensional 
\lq internal space' which is non-compact. Thus the supergravity theory suggests
that the way round the no-go theorem of [\gib,\mn] is to have such a \lq
non-compactification'. In general,  solutions of this kind with non-compact extra
dimensions  will not have an effective description in terms of a $d$ dimensional
theory and will be intrinsically $D$ dimensional. However, it is possible that in
some cases such solutions will lead to $d$-dimensional physics. Examples with
large extra dimensions in which this is so are provided by [\randsum]. In the
present context, one approach is to impose suitable boundary conditions on the
\lq internal' non-compact space so that there is a mass gap and a sensible $d$
dimensional spectrum with light fields and infinite towers of massive states
[\gel,\CW,\NicWet].  A useful criterion is to seek solutions in which there is a
consistent truncation to a $d$-dimensional supergravity theory, and the ones
obtained by \lq lifting' a $d$ dimensional theory to $D$ dimensions are usually
of this kind.

It is perhaps worth recalling that when gauged supergravities were first found,
it was   thought that they were disappointingly unphysical as they had
supersymmetric anti-de Sitter vacua which  appeared to be unstable. For example,
the gauged $N=4$ supergravity in $D=4$ of [\nfor] has a complex scalar
$\phi$ with  potential
$$V=-\2 g^2(cosh (a \vert  \phi  \vert ) +2)
\eqn\potl$$ where $g$ is the gauge coupling constant and $a$ is a constant,
which could be absorbed into the definition of $\phi$. This has a maximum at
$\phi=0$ leading to a spacetime with negative cosmological constant
$$\lll=-{3\over 2}g^2
\eqn\abc$$ and flat-space intuition suggested that a theory with such a scalar
potential should be unstable. However, the anti-de Sitter vacuum was later shown
to be completely  stable [\adsstab] and more recently the physical relevance of
anti-de Sitter vacua has become understood [\mal].

There are also gauged supergravities with de Sitter vacua, and  these again
appear unstable as the critical point is a maximum. For example, the  gauged
$N=4$ supergravity in $D=4$ of [\gat] again 
has a complex  scalar
$\phi $ with  potential
$$V=-\2 g^2(cosh (a   \vert \phi   \vert ) -2)
\eqn\pota$$ with a maximum at $\phi=0$ leading to a spacetime with positive
cosmological constant
$$\lll=\2 g^2
\eqn\abc$$ and there are similar de Sitter vacua of gauged $N=8$ supergravities in
$D=4$ [\nct] and $D=5$ [\gunwar] (although none are known in $D=7$). These were
lifted to  solutions of suprgravity in 10 or 11 dimensions in [\CW].
 Such de Sitter vacua break all supersymmetries, as was to be expected, and while
such an upside-down potential is less unstable in anti de Sitter space than in
flat space, due to the Breitenlohner-Freedman mechanism, it is liable to be more
unstable in de Sitter space. Nonetheless, it is remarkable that such a structure
is forced by supersymmetry, and it would be worth investigating whether such
theories could be of relevance. For example, there could be a long period
of inflation in which the scalar field  rolls slowly down the  potential, after
which the vacuum might decay to a different  (and perhaps   realistic) solution. 
 These solutions will be discussed in section 2.

As well as these de Sitter solutions of conventional supergravities, de Sitter
space also arises as solutions of certain \lq variant' supergravity theories
[\vans,\tim,\time].  There are de Sitter superalgebras [\vans,\luk] -- typically
they are different real forms of the more familiar anti-de Sitter superalgebras --
but they do not have unitary highest weight representations. There are field
theory realisations of these  as supergravity theories in which  the lack of
unitary representations is reflected in the  fact that some of the fields have
kinetic terms with the wrong sign. These have de Sitter solutions which are
invariant under the full de Sitter supergroup, but the lack of unitarity makes
these problematic field theories. However, these arise from string theories which
are formally related to the conventional string theories by dualities (such as
T-dualities in a compact time direction) suggesting that these theories might be
worth re-examining in a string theory context. 
These arise as 10 or 11 dimensional solutions which are a direct product of de
Sitter space with   a non-compact hyperbolic space, but in these cases 
the \lq internal space' can be compactified by identifying under the action of a
discrete isometry group. (This is not the case for 
the de Sitter solutions of conventional supergravities described above, as the \lq
internal space' does not have a suitable discrete isometry group that can be used
for such an identification.)  These variant supergravities do not satisfy
the conditions assumed in the
  theorems of [\gib,\mn].
 One advantage of these theories is
that the supersymmetry of the de Sitter background can give clues as to how to
treat de Sitter space that might be more widely applicable. These will be
discussed in sections 3 and 4.

In [\tim], it was argued that de Sitter space supergravity or string theory in $D$
dimensions should have a holographic dual which is a  $D-1$ dimensional 
Euclidean  conformal field theory, with the 
de Sitter group $SO(D,1)$ acting as the conformal group in $D-1$
dimensional Euclidean space.
This arose from an argument similar to that used in [\mal] for  the anti-de Sitter
case. The  variant supergravities have brane-like
solutions that interpolate between flat space and the de Sitter solution, 
which arises as a near-horizon limit.
However
these branes
are spacelike surfaces with effective worldvolume theories that are Euclidean field
theories, instead of the timelike surfaces of conventional branes which have 
Lorentzian worldvolume field theories. Following [\mal], it was argued in
[\tim] that the string theory in the de Sitter space arising in the near-horizon
limit should provide a dual description of the Euclidean world-volume field theory,
with  the de Sitter suprgravity 
arising as a 't Hooft limit of the field theory. These variant supersymmetric
theories appear to be non-unitary, but the results of [\strom] suggest that
non-unitarity 
could be a typical feature of de Sitter holography.
  As in the anti-de Sitter case, such a
duality arises in more general contexts and can be extended to   cases with
less supersymmetry. This will be discussed further in sections 7-9.

The presence of a cosmological horizon in de Sitter space raises a number of issues
as to how to quantise in a de Sitter vacuum, one   being whether the
degrees of freedom to be used should be those in a causally connected region
[\banksb] or ones in the whole space [\witten].
It is perhaps worth mentioning that  tachyons, such  as those that arise   
in some of
the theories  
   discussed above, can escape through a horizon
and for such theories it would not be appropriate  to limit considerations to
degrees of freedom within the horizon.

Finally, the construction of [\hlw] or [\pope] provides another class of variant
supergravities with de Sitter solutions [\pope,\chamlam]; these will not be
discussed here.

\chapter{De Sitter Space from Gauged Supergravity}

In $D=4$ the Cremmer-Julia $N=8$ supergravity theory [\CJ], with scalars in the
coset space $E_7/SU(8)$,  can be gauged by promoting a subgroup of the rigid
$E_7$ symmetry to a local symmetry. The $SO(8)$ gauging of [\dwn] has a maximally
supersymmetric anti-de Sitter vacuum and can be truncated to the gauged $N=4$ theory
with potential \potl. In [\nct], gaugings with gauge group
$CSO(p,q,r)$ were obtained for all non-negative integers $p,q,r$ with $p+q+r=8$,
where
 $CSO(p,q,r)$ is 
the group contraction of $SO(p+r,q)$
 preserving a symmetric metric with $p$ positive eigenvalues, $q$
negative ones and $r$ zero eigenvalues. Then  $CSO(p,q,0)=SO(p,q)$ and
$CSO(p,q,1)=ISO(p,q)$. In [\frefo], it was argued that these are the only
possible gauge groups. Note that despite the non-compact gauge groups, these are
unitary theories, as the vector kinetic term is not the minimal term
$k_{ab} F^a \cdot F^b$ contracted with the indefinite Cartan-Killing metric  
$k_{ab}$, but is   $Q_{ab}(\phi) F^a \cdot F^b$ contracted
with a positive definite scalar-dependent matrix  $Q_{ab}(\phi)$.
Of these theories, the ones with gauge groups
$SO(4,4)
$ and
$SO(5,3)$ have de Sitter vacua arising at local maxima of the potentials, and the
$SO(4,4)
$ theory includes the gauged $N=4$ theory with potential \pota\ as a sub-sector
[\nct]. In $D=5$, the gauged $N=8$ supergravities include those with gauge groups
$SO(p,6-p)$ [\gunwar,\PPV] and of these the $SO(3,3)$ gauged theory has a de
Sitter vacuum.

In each of these cases, the de Sitter vacua break all supersymmetries and break
the gauge group $SO(p,q)$ down to its maximal compact subgroup
$SO(p)\times SO(q)$. The higher-dimensional origin of these theories was found in
[\CW]. The gauged supergravities in $D=4$ with gauge group  $CSO(p,q,r)$ arise
from warped configurations of 11-dimensional supergravity with
\lq internal' space of the form
$\H ^{p,q,r}$ where $\H^{p,q,r}$ is the hypersurface of $\R^{p+q+r}$ in which the
real Cartesian coordinates $z^A$ of 
$\R^{p+q+r}$ satisfy
$$\eta _{AB}z^A z^B=R^2
\eqn\abc$$ Here $R$ is a constant \lq radius', and $\eta _{AB}$ is a constant
metric with  $p$ positive eigenvalues, $q$ negative ones and $r$ zero eigenvalues.
The metric on the hypersurface $\H^{p,q,r}$ is the positive definite metric induced
from the Euclidean metric on
 $\R^{p+q+r}$.
 Thus   $\H ^{p,0,0}$ is a sphere $S^{p-1}$, 
$\H ^{p,1,0}$ is the hyperboloid $H^p$, which is the coset space $SO(p,1)/SO(p)$,
$\H ^{p,q,0}$ is a hyperbolic
space (a non-symmetric space with negative curvature)
 and $\H ^{p,q,r}=\H ^{p,q,0}\times \R^r$ is a
generalised cylinder with cross-section $\H ^{p,q,0}$. For the cylinders, the
flat directions can be compactified to give
$\H ^{p,q,0}\times T^r$. 

If $d\www ^2_{p,q,r}$ is the metric on $\H ^{p,q,r}$
induced from the  Euclidean metric on $\R^{p+q+r}$, then the solutions
studied in [\CW] include those with $r=0$ and warped product metrics in
$D=d+p+q-1$ dimensions of the form 
$$ds^2=f_1^2(y) dS_d^2(x) + f_2^2(y)d\www ^2_{p,q,0}(y)
\eqn\abc$$ 
 where $dS_d^2(x)$ is the metric of a solution of the $d$-dimensional
gauged supergravity   theory  with coordinates $x$, while  $y$ are intrinsic
coordinates on $\H ^{p,q,0}$. Here  $$\eta_{AB}=diag (\II_p,-c^2 \II_q)
\eqn\abc$$ for some constant $c$ and the warp factors are given in terms of
$$L^2 =R^{-2}\left[ \sum _{i=1}^p (z^i)^2 +c^4 \sum _{i=p+1}^{p+q} (z^i)^2\right]
\eqn\abc$$ by
$$f_1=L^a, \qq f_2=L^b
\eqn\abc$$ for some constants $a,b$. As in the Freund-Rubin ansatz, there is
an antisymmetric tensor field strength that is proportional to the volume form
of one of the two factors.
The cases of interest here are the ones in which $dS_d^2$ is the metric on
$d$-dimensional de Sitter space.  In $d=4$ the de Sitter solution of the
$SO(4,4)$ gauged theory arises from the solution of 11-dimensional
supergravity of this form with $$p=4,\qq q=4,\qq c^2=1,\qq a=2/3, \qq b=-1/3
\eqn\abc$$ while the de Sitter solution of the
  $SO(5,3)$ gauged theory arises from the solution of 11-dimensional supergravity
of this form with
$$p=5,\qq q=3,\qq c^2=3,\qq a=2/3, \qq b=-1/3
\eqn\abc$$ The de Sitter solution of the $SO(3,3)$ gauged theory in $d=5$
arises from a similar solution of IIB  supergravity with $d=5$ and
$$p=3,\qq q=3,\qq c^2=1,\qq a=1/2, \qq b=-1/2
\eqn\abc$$ and with the self-dual 5-form field strength given in terms of the
volume forms on $dS_5$ and the internal space.

These solutions can be found by an analytic continuation of the $S^7$
compactification of 11-dimensional supergravity and the
$S^5$ compactification of IIB supergravity [\CW]. The sphere reductions have
consistent truncations to the gauged supergravity sector, and it follows from the
structure of the analytic continuation that the \lq non-compactifications' from
11 dimensions on
$\H ^{4,4}$ or $\H ^{5,3}$ or from IIB supergravity on $\H ^{3,3}$ have  
consistent truncations to the corresponding gauged supergravities (similar
arguments  were used  in [\popec]).

\chapter{De Sitter Supergravities}

There is a class of variant supergravities which have maximally supersymmetric de
Sitter vacua invariant  under a de Sitter supergroup. By invariance, it is meant
that the $d$-dimensional classical  solution is invariant under the  de Sitter
isometry group $SO(d,1)$, the supersymmetry transformations generated by    a
(maximal) set of Killing spinors, and an R-symmetry group which is typically
non-compact. (The definition of corresponding conserved charges is problematic
in de Sitter space.)  The supergravity theories in general have some fields
with kinetic terms with the wrong sign, as is needed for invariance under a
linearly-realised non-compact R-symmetry. These theories are typically
analytic continuations to different real forms of the gauged supergravities
that give anti-de Sitter space. The first such theory, constructed in [\vans],
was a variant form of   $N=2$ gauged supergravity. The usual gauged $D=4,N=2$
theory has gauge group $U(1)$ with charged gravitini and a negative
cosmological constant. In the variant form of [\vans], the sign of the vector
field kinetic term is reversed and the theory has a positive cosmological
constant.

Maximally supersymmetric generalisations of this with 32 supersymmetries were
found in [\tim,\time]. The  usual ungauged $N=8$ supergravity in $d=5$ has
scalars in the coset
$E_{6(+6)}/USp(8)$ and in [\tim] a variant form (again in 4+1 dimensions and
with 32 local supersymmetries) was found with scalar coset structure
$E_{6(+6)}/USp(4,4)$; this will be referred to here as the $N=8^*$ theory. As
$USp(4,4) $ is non-compact, some of the scalar fields have kinetic terms of
the wrong sign, as do many of the other matter fields. There are gauged
versions of this $N=8^*$   theory with gauge groups $SO(p,6-p)$ and in
particular that with gauge group $SO(5,1)$ has a $d=5$ de Sitter vacuum
invariant under the de Sitter group $SU^*(4/4)$ with bosonic subgroup
$SO(5,1)\times SO(5,1)=SU^*(4)\times SU^*(4)$ and  with 32 fermionic
generators, corresponding to the 32 Killing spinors [\tim]. The other
$SO(p,6-p)$  gauge groups can be obtained from the $SO(5,1)$ gauging using the
methods of [\nct]. 

Similarly, in 3+1 dimensions, there is a variant $N=8^*$ supergravity with
coset structure $E_{7(+7)}/SU(4,4)$ [\tim] instead of the structure
$E_{7(+7)}/SU(8)$ of the Cremmer-Julia theory, and this has a gauging with
gauge group $SO(6,2)$ [\time]. This has a de Sitter vacuum invariant under the
super group $OSp^*(4/8)$, with bosonic subgroup
$SO(4,1)\times SO(6,2)$ and 32 supersymmetries.  $N=8^*$ theories with gauge
groups $CSO(p,q,8-p-q)$ can be obtained from this by the methods of [\nct].   

Such variant theories cannot arise from dimensional reduction of conventional
supergravities, and their higher dimensional origin must be from
  variant supergravities. The $IIA^*$ and $IIB^*$ supergravities in 9+1
dimensions [\tim] are variant forms of the usual type IIA and IIB
supergravities  in which  the kinetic terms of the Ramond-Ramond gauge fields
all have the wrong sign. The bosonic actions are
$$\eqalign{S_{IIA^*}= 
\int
 d^{10} x &
\sqrt{-g}\left[
 e^{-2 \Phi} \left(  R+ 4(\partial   \Phi )^2   -H^2 
\right)\right.
\cr &
\left. +G_2^2 +G_4^2 
\right] + \dots
\cr}
\eqn\twoas$$ and
$$
\eqalign{S_{IIB^*}= 
\int
 d^{10} x &
\sqrt{-g}\left[
 e^{-2 \Phi} \left( R+ 4(\partial   \Phi )^2   - H^2 
\right)\right.
\cr &
\left. +G_1 ^2+  G_3 ^2+ G_5 ^2 \right] + \dots
\cr}
\eqn\twobs$$ where the field equations  from \twobs\ supplemented by the
constraint $G_5=*G_5$. The scalars in the $IIB^*$ theory take values in the coset
$SL(2,\R)/SO(1,1)$, with the Ramond-Ramond scalar having a  kinetic term with the
reversed sign. 
This sign reversal leads to brane solutions that carry Ramond-Ramond charge being
spacelike (E-branes) rather than timelike (D-branes).

Reducing either of the type $II^*$ theories on a (Euclidean) 5-torus gives the
  ungauged $N=8^*$ supergravity in 4+1 dimensions  while reducing on a
6-torus gives the   ungauged $N=8^*$ supergravity in 3+1 dimensions.
 The five dimensional $N=8^*$ supergravity
 theory with gauge group $SO(5,1)$ is obtained as a consistent
truncation of the $IIB^*$ theory in the solution $dS_5\times H^5$ [\tim],
where $H^d$ is the hyperbolic space 
$$H^{d}={SO(d,1)\over SO(d)}
\eqn\abc$$ with isometry group $SO(d,1)$, while
$d$ dimensional de Sitter and  anti-de Sitter spaces are the coset spaces
$$dS_d={SO(d,1)\over SO(d-1,1)},\qq AdS_d={SO(d-1,2)\over SO(d-1,1)}
\eqn\abc$$ with isometry groups $SO(d,1)$ and $SO(d-1,2)$ respectively.

The $IIA^*$ theory cannot be obtained from any variant theory in 10+1
dimensions, but   arises from  compactifying a supergravity theory in 9+2
dimensions on one of the timelike dimensions [\time]. The de Sitter theory in
4  dimensions can be obtained from a solution of this $M^*$ supergravity 
given by the product   $dS_4\times AdS_7$, with the gauge group $SO(6,2)$
arising as the isometry group of $AdS_7$. Regarding the $AdS_7$ as the
internal space, there is a consistent truncation to the 4-dimensional variant
gauged supergravity.

The dimensional reductions considered in this section are all analytic
continuations of the  sphere reductions of conventional supergravities, and so
the consistency of the truncations of these to lower dimensional supergravity
theories implies the consistency of the reduction here also. Note that only
certain analytic continuations can be consistent with supersymmetry, and for
example 7-dimensional de Sitter space does not arise [\time,\HK,\HKt]; the
possible analytic continuations of the $AdS_7\times S^4$ solutions of
11-dimensional supergravity that do arise in this way are given in [\HK,\HKt].

\chapter{String Theory, Time and Duality}

The solutions of 11-dimensional supergravity and IIB supergravity of section 2
are classical solutions of M-theory or IIB string theory, although the tachyonic
scalar potential appears to  signal an instability. The de Sitter supergravities
and their origin in $IIB^*$ supergravity or
$M^*$ theory are more problematic,  as the field theory limits have terms in the
supergravity lagrangian with the \lq wrong' sign, but the arguments of [\tim,\time]
 suggest that there is a formal link  to the usual M-theory via dualities
involving the time dimension, and these will now be reviewed, as they provide a
formal link between the AdS supergravities and the de Sitter ones, and motivate
the holography conjectures to be discussed in later sections.

Consider type IIA or IIB string theory in flat space-time but with time
periodically identified, with  $t\sim t+2\pi R$. If the periodicity is extremely
large  then  one might expect the physics to be  similar to that in Minkowski
space.
 Such a background is certainly a  solution of the theory, but issues arise as to
whether the quantum theory makes sense with periodic time. Quantum mechanics or
quantum field theory with periodic time has a number of unusual features. With
periodic boundary conditions in time, one can solve the Schr\" odinger or wave
equations, giving quantised frequencies,  and one can perform the functional
integral and  calculate quantum correlation functions, but the interpretation is
problematic. There is not  a conventional  probabality interpretation in such
circumstances, as the result of any \lq experiment' would be determined by what
happened last time round, and the wave function would have already collapsed.
However, the problems of quantum interpretation are similar to those that arise
in addressing the quantum behaviour of the whole universe in a cosmological setting.
Indeed, in  a periodic cosmology  in which the universe expands and then
contracts and   then repeats the cycle, so that time is periodic with a  
recurrence time given by 
  the \lq lifetime' of the universe, the problems of interpretation become
the same as those of quantum cosmology. There are many suggestions in the
literature as to how such  issues can be addressed, but whatever the resolution,
there should be some  description of the universe as a whole which is quantum in
nature.  String theory can presumably be formulated in a cosmological spacetime
or in a spacetime with periodic time, and in both cases     issues of
interpretation arise.

If time is identified with radius $R$, energy is quantised in units of $1/R$ and it
is sometimes suggested that this could be in conflict with 
the quantization of mass in string theory in units of the string mass $m_s$, unless
the radius $R$ were related to $1/m_s$. This is not so; the energy $E=p^0$ is
quantised
$$E={2\pi n \over R}
\eqn\enqu$$
for some integer $n$
and the string physical state conditions for a state with momentum
$p^\mm=(E,\bf p)$ give
$$ E^2 - {\bf p} ^2 = m_s^2 N
\eqn\mqu$$
where $N$ is an integer given in terms of the eigenvalues of   the oscillator
number operators. These two conditions would clearly be in conflict in general if
the spatial momentum  ${\bf p}$ vanished, but compactifying time breaks Lorentz
invariance and one can no longer use a Lorentz transformation to go to the rest
frame ${\bf p}=0$.
The two conditions are compatible with  ${\bf p}\ne 0$, and for 
any given $N$, one can find an energy satisfying \enqu\ and a momentum ${\bf
p}$
such that \mqu\ is satisfied, but ${\bf
p}$ will be non-zero if $m_s^2R^2/4\pi ^2$ is irrational.

If string theory exists with periodic time, its properties can be analysed by
standard methods, and in particular one can perform a T-duality in the time
direction. In the functional integral, time is Wick rotated to $\tt=-it$, and the
Euclideanised functional integral with a periodic coordinate will exhibit
T-duality. However, there are a number of different ways  of
continuing the Euclideanised theory back to Lorentzian signature, depending on
whether the periodic coordinate is continued back to a spacelike or a timelike
coordinate, or is treated as the periodic Euclidean time corresponding to a finite
temperature. In the case at hand, a   timelike circle of radius $R$ is T-dual to
a timelike circle  of radius $R'=  \aa'/R$. Such a timelike T-duality takes the
bosonic string to the bosonic string and the heterotic string to the heterotic
string, but takes type IIA or IIB string theories to new theories, the $IIB^*$
and $IIA^*$ theories whose field theory limits are the  $IIB^*$ and $IIA^*$
supergravity theories [\tim]. For spacelike T-duality, IIA string theory on a
circle of radius
$R$ is T-dual to
 IIB string theory on a circle of radius
$R'=  \aa'/R$. For  timelike T-duality, IIA (IIB) string theory on a timelike
circle of radius
$R$ is T-dual to
$IIB^*$ ($IIA^*$) string theory on a circle of radius
$R'=  \aa'/R$ [\tim]. The \lq wrong' signs of 
 the kinetic terms of the RR fields can be understood from matching the
dimensional reductions of the supergravities on a timelike circle. For example,
dimensional reduction of IIA supergravity on a timelike circle to 9 Euclidean 
dimensions
 gives a RR scalar $C=C_0$ and 2-form $C_{ij}=C_{0ij}$ whose kinetic terms will
be of the \lq wrong' sign, and if they are to come from the reduction of a RR
scalar field and 2-form field in a dual IIB-like theory in 9+1 dimensions, these
fields must have kinetic terms of the \lq wrong' sign.

The $IIB^*$ ($IIA^*$) string theory on a background with a timelike circle of
radius $R$ is then precisely the IIA (IIB) string theory on the dual timelike
circle with radius $R'=  \aa'/R$, but written in terms of different variables.
Either both theories exist, or  neither do; if both exist, then the problems with
one can be translated to the problems of the other via the duality. With periodic
time, the \lq wrong' signs may not be as bad as they at first appear, as
instability or loss of unitarity are not the problems that they would be in flat
space. A theory which would be classically unstable in flat space, due for
example to terms of the wrong sign in the action leading to an energy density
which is not positive, would not be unstable with periodic time: any instability
which started to grow would have to shrink again to satisfy the periodic time
boundary conditions. Similarly,  non-unitarity is often associated with loss of
\lq probability', but again the boundary conditions would  result in any \lq
probability' that is lost having to come back again. There is a similar escape
clause for many of the problems usually associated with \lq wrong' signs, and it
is conceivable that the theory could be consistent in a background with compact
time. In the decompactification limit $R'\to \infty$, the theory in Minkowski
space appears to be unstable, due to the non-positive kinetic energy.

This suggests that while the type $II^*$ theories in Minkowski space could be 
unstable (or worse), they could be better behaved in backgrounds with compact
time. If so, it is also possible that other backgrounds of the type $II^*$
theories could be viable, and in particular the theory in de Sitter or
comsological backgrounds may not  be as bad as they at first appear. For example,
as will be discussed in section 8, there are  supersymmetric cosmological
solutions  with an expanding time-dependent geometry   in which a RR gauge field
grows with time, with a time dependence  given in terms of $H(t)$, and it is not
clear whether a solution  in which a field   grows as the universe expands should
be viewed as an instability.
 (Some  solutions of this type are obtained from   an analytic continuation of
brane solutions, in which  a radial coordinate $r$ is replaced by a time
coordinate $t$.) The properties  of the theory in such cosmological backgrounds
deserve
  further study.

If the $IIA^*$ theory in flat space needs a compact time, then $M^*$ theory in
flat space in 9+2 dimensions would need both timelike dimensions to be periodic.
Dimensionally reducing $M^*$ theory from 9+2 dimensions on a spacelike circle
gives a IIA-like theory in a 
 spacetime in 8+2 dimensions, and further dualities generate type II string
theories in all spacetime signatures $(s,t)$ with $s+t=10$, and leads to one
further 11-dimensional theory, the $M'$ theory in 6+5 dimensions [\time]. In each
case, there is a   supergravity theory, although the details are different in
each case, as the properties of spinors are  sensitive to the signature [\time].
If type II string theory   exists  in  flat space with periodic time, then
these other type II string theories and the $M^*,M'$ theories should also exist
and are related to  
$M$-theory by chains of dualities. In this way, the de Sitter  solutions of the $
IIB^*$ or $M^*$ theories  could be regarded as solutions of exotic phases of
M-theory.

\chapter{Non-Compactifications}

De Sitter space can be obtained in supergravity theories  
in higher dimensions if 
  the extra dimensions take the form of a non-compact hyberbolic space, and the
no-go theorem of [\gib,\mn] suggests that this will be generic, and that de Sitter
spaces in M-theory will typically be accompanied by a non-compact internal space,
motivating a reconsideration of such spaces as solutions. In the examples
considered here, there is a consistent truncation to  a lower dimensional
supergravity theory, arising from configurations in which fields in the internal
space are in their ground states.
 The question then arises  as to whether this can be extended to allow general
configurations with non-trivial dependence on the internal space   while still
being able to  extract sensible lower-dimensional physics.

In some cases, one can compactify a non-compact internal space. If the internal
space is $\R^n$, one can identify points under a discrete subgroup of the
translation group to obtain a torus $T^n$. Less trivially, for the hyperbolic
space $H^d$, one can identify points under the action of a discrete subgroup of
the isometry group $SO(d,1)$ to obtain a compact space $\ti H^d$ and consider a
conventional compactification on $\ti H^d$. Thus the solution
$dS_5\times H^5$ of the $IIB^*$ theory can be replaced by the compactifying 
solution
$dS_5\times \ti H^5$.  With such a compact internal space, one can dimensionally
reduce in the standard way. However, this will not work for the spaces
$\H^{p,q,0}$ with isometry group
$SO(p)\times SO(q)$ as they do not have a non-compact
\lq translational' isometry group that can be used in such an identification.

For a non-compact internal space,  a standard dimensional reduction of the action
leads to a reduced theory with Newton's constant inversely proportional to the
volume of the internal space, so that it would vanish for an infinite volume
internal space. However, for the classical theory, it is sufficient that the
dimensionally reduced field equations make sense and this is possible even for an
infinite volume internal space.  
In many cases,  boundary
conditions can be imposed so as to obtain a discrete spectrum.
 Consider for
example  a theory on $M\times N$, where $N$ is
regarded as the
\lq internal' space and consider  a scalar field $\Phi(x,y)$  satisfying
a wave equation
$$\ddd  \Phi =- m^2\Phi\eqn\abc$$ where $\ddd$ is  the   Laplace
operator on $M\times N$, $x$ are the coordinates on $M$ and $y$ those on $N$, and
the Laplacian splits into Laplacians acting on $M$ and $N$,
$\ddd=\ddd_M+\ddd_N$. 
It was shown in [\NicWet] that in many cases the spectrum of the Laplace operator
$\ddd_N$ consists of a discrete spectrum of normalisable modes and a continuous
spectrum of non-normalisable ones.
If boundary conditions can be imposed on $N$ that eliminate the non-normalisable
modes, one is left with a
discrete set of eigenfunctions $f_n(y)$ of the Laplacian $\ddd_N$ satisfying
$$\ddd_N f_n=\ll_n f_n\eqn\abc$$ In such cases, a
 Kaluza-Klein-type  spectrum emerges.
A similar analysis extends to higher spins  and to   warped products
[\NicWet].

Consider next the interactions. Suppose  there is set of scalar fields
$\Phi ^a$ on $M\times N$ labelled by $a$      with field equations
$$
\ddd\Phi^a=-M^a{}_b \Phi^b+ c^a_{bc}\Phi^b\Phi^c+O(\Phi^3)
\eqn\abc$$ with mass matrix $M^a{}_b$  and coupling constants  $c^a_{bc}$. Then
the decomposition
$$\Phi^a(x,y)=\sum _n \phi_n^a(x) f_n\eqn\abc$$ will lead to well-defined  field
equations on $M$ provided the eigenfunctions satisfy a completeness relation
$$f_m(y)f_n(y)=\sum _p d^p_{mn}f_p(y)
\eqn\abc$$ for some finite constants $d^p_{mn}$. Then the resulting field
equations on $M$ are
$$\ddd_M \phi_n^a=  M^a{}_b\phi_n^b- \ll_n\phi_n^a
+c^a_{bc}d^n_{pq}\phi^b_p\phi^c_q +O(\phi^3)
\eqn\abc$$ This extends to other interactions and general spins, with the result
that the field equations on  $M\times N$ can be reduced to well-defined  field
equations on $M$, much as in a compactification, provided  boundary conditions
are imposed on the fields on $N$ such that the wave operators on $N$ all have
discrete spectra and in addition the eigenfunctions satisfy   appropriate
completeness conditions.


\chapter {Holography in Product Spaces}

Supergravity or string theories in a space $M$ can have a holographic description
in terms of a theory on the boundary of $M$, $\pa M$. In [\HKt], this was
generalised to theories on a product   space
$M\times N$ in which both $M$ and $N$ had boundaries, and it was argued that two
dual field theories can  play a role here, one on $\pa M$ and one on
$\pa N$.  The boundary of $M\times N$ has two components,
$\pa M \times N$ and $\pa N \times M$, and in general there could be holographic
dual theories on   these two boundaries. Suppose $N$ is compact, or that boundary
conditions are applied so that there is a discrete spectrum and the  theory on
$M\times N$  can be regarded as a theory on $M$ with an infinite tower of massive
fields, as discussed in the last section. Then there could be a dual description
as a theory on the boundary     $\pa M $ of $M$ (and similarly with the roles of
$M$ and $N$ interchanged). 

For theories on  a supersymmetric $AdS_d\times S^n$ background, there is a
holographic description in   terms of a $d-1$-dimensional conformal field theory
(CFT), which can be thought of as being on the
$d-1$-dimensional boundary of $AdS_{d}$ [\mal], with the AdS isometry group
$SO(d-1,2)$ acting as  the conformal group on the boundary. There is a
correspondence between boundary values of fields in anti-de Sitter space and
operators in the conformal field theory, and in particular the Kaluza-Klein modes
representing fields with non-trivial dependence on the internal dimensions
correspond to certain operators in the CFT.

The correspondence can be formulated in terms of the Wick rotated theory, in
which 
    the $AdS_{d}$ space is analytically continued to the $d$ dimensional
hyperbolic space 
$H^{d}=SO(d,1)/SO(d)$ and the boundary   theory is continued to a
Euclideanised conformal field theory on the boundary $S^{d-1}$ of $H^d$ [\witt].
As in [\tim], it will be convenient to refer to supersymmetric field theories 
formulated directly in Euclidean space as \lq Euclidean', and to the theories
obtained by Wick rotating supersymmetric theories from Lorentzian space as \lq
Euclideanised';  the Euclideanised theories will usually not have a conventional
supersymmetry.
For example, Wick rotating Lorentzian $N=4, D=4$ \sym\ gives a Euclideanised
theory in 4+0 dimensions with $SO(6)$ R-symmetry and no conventional
supersymmetry.
A Euclidean \sym\ theory with 16 supersymmetries in 4+0 dimensions and $SO(5,1)$
R-symmetry is obtained by reducing \sym\ from 9+1 dimensions on 5 space and one
time dimension [\Bob,\thom].

A similar situation applies for other products of non-compact spaces, when each
has a holographic dual. Consider the product $dS_n\times H^d$, with isometry group
$SO(n,1)\times SO(d,1)$.  Both de Sitter space and 
 hyperbolic space are non-compact, and it was argued in [\tim,\HKt] that there
should be  two holographic descriptions arising from different limits, one
which is a $d-1$ dimensional Euclidean field theory on the boundary of
$H^d$ and one  which is an $n-1$ dimensional Euclidean field
theory associated with de Sitter space. The space $dS_n\times H^d$ can be
Wick rotated to the Euclideanised solution 
$S^n\times H^d$, which is exactly the same as the Euclideanisation of
$AdS_d\times S^ n$. This Euclideanised theory   has a holographic  description as
a theory on the boundary of $H^d$, which continues back to a holographic
description of the $dS_n\times H^d$ theory on the boundary of $H^d$. Here
$SO(d,1)$ acts as the conformal group on the Euclidean $d$-dimensional CFT while
$SO(n,1)$ arises as the R-symmetry group. As de Sitter space becomes a sphere on
Euclideanisation, the Euclideanised theory does not help in formulating the de
Sitter holography. In [\tim] it was argued  that in certain circumstances, to be
discussed in the following sections,   the physics in $n+1$ dimensional de
Sitter space has a holographic description as an $n$-dimensional  Euclidean
conformal field theory with  conformal group $SO(n,1)$ and    R-symmetry
$SO(d,1)$.
Thus there are two holographic duals: one on the boundary of $H^d$ with conformnal
group $SO(d,1)$ and one associated with  the boundary of de Sitter space
with conformal group $SO(n,1)$.

For example,   the $AdS_5\times S^5$ solution of IIB string theory has a 
holographic representation as $D=4,N=4$ super Yang-Mills theory on the boundary
of $AdS_5$, with super-AdS group $SU(2,2/4)$ realised as the superconformal group,
with conformal group $SO(4,2)$ and R-symmetry group $SO(6)$.
  Wick rotation takes this to  a Euclideanised theory on $H^5\times S^5$ with
isometry group
$SO(5,1)\times SO(6)$, and the holographic dual is the Euclideanised
super-Yang-Mills theory  with R-symmetry $SO(6)$,  the $SO(3,1)$ Lorentz symmetry
continued to $SO(4)$ and the conformal group continued to $SO(5,1)$. Neither
theory has conventional supersymmetry, due to the usual problems in continuing
spinors, self-dual forms and  supersymmetries to Euclidean space. 
(Supersymmetric theories can be obtained after further sign changes, so that the
R-symmetry becomes $SO(5,1)$, to give 
Euclidean supersymmetric   theories.)

The $dS_5\times H^5$ solution of the $IIB^*$ theory
 is invariant under the super-de Sitter group $SU^*(4/4)$, which contains the
isometry group
$SO(5,1)\times SO(5,1)$. This has a holographic dual description [\tim] in terms
of a Euclidean superconformal Yang-Mills theory in four Euclidean dimensions
obtained by reducing super-Yang-Mills from 9+1 dimensions on one time and  five
space    dimensions [\Bob,\thom]. This has conformal group $SO(5,1)$ and
R-symmetry group
$SO(5,1)$. Here $n=d=5$, but the Euclidean conformal field theory can arise in
two ways, one on the boundary of $H^5$ and one associated with the de Sitter
space, as will be reviewed in section 9. The Euclideanised version of this theory
is the same theory on $H^5\times S^5$ as obtained from continuing the
$AdS_5\times S^5$ solution. The Euclidean super-Yang-Mills theory has 5 scalars
with kinetic terms of the \lq right' sign and one of the \lq wrong' one, so that
the $SO(5,1)$ R-symmetry can be linearly realised on them, and the
Euclideanisation involves multiplying the wrong-sign scalar by $i$ to get a
positive action and $SO(6)$ R-symmetry (just as in the string theory path
integral, one continues $X^0\to i X^0$).

\chapter {Euclidean Branes}

T-duality  exchanges Dirichlet and Neumann string boundary conditions. A
Dirichlet $p$-brane  at $y^i=0$ in   flat spacetime with coordinates
$X^M=(t,x^1,\dots x^p,y^1,\dots y^{9-p})$ 
  corresponds to  strings $X^M(\ss,\tt)$ with Neumann boundary conditions on the  
$p+1$ longitudinal coordinates $t,x^i$ and Dirichlet boundary conditions on the
transverse coordinates $y^i$. A T-duality in a particular direction changes the
boundary conditions in that direction from Dirichlet to Neumann or vice versa,
so   a T-duality in a longitudinal spatial  direction $x^p$ takes it to a $p-1$
brane with
$p$ longitudinal coordinates $t,x^1,...,x^{p-1}$.
If the time
direction is taken to be compact,  a  T-duality
in the time direction takes a type II theory to a type $II^*$
theory and changes the boundary conditions on $t(\ss,\tt)$
from Neumann to Dirichlet, giving a $p$-dimensional spacelike surface parameterised
by the  coordinates $x^1,...,x^{p }$ located at $y^i=0$ and at a fixed moment in
time, $t=t_0$ for some constant $t_0$.
This is the E$p$-brane of the type $II^*$ theory [\tim].

The world-volume theory on a stack of $N$ D$p$-branes is the $p+1$ dimensional 
\sym\
theory with gauge group $U(N)$ and R-symmetry $SO(9-p)$ obtained by reducing 9+1
dimensional
\sym\ on a
$9-p$ torus. For a single D-brane, there are $9-p$ scalar fields which are
collective coordinates representing the position of the brane; changing their
expectation values changes the position of the brane in the Euclidean transverse
space.
For $N$
E$p$-branes, 
the world-volume theory  is the $p $ dimensional Euclidean
\sym\
theory with gauge group $U(N)$ and non-compact R-symmetry $SO(9-p,1)$ obtained by
reducing 9+1 dimensional \sym\ on a Lorentzian  torus with $9-p$ spacelike circles
and one timelike circle.
Again the expectation values of the $10-p$ scalars correspond to the position in the
transverse space, but here the transverse space is Lorentzian.
This is reflected in the fact that $9-p$ of the scalars have the conventional
sign of kinetic term, representing the spatial position of the brane, and one has
the wrong sign kinetic term and changing its expectation value changes the brane
instant $t=t_0$.
For $p=4$, this gives the  conformally invariant \sym\ theory in four Euclidean
dimensions.

\chapter{Interpolating Solutions and Branes}

For a wide class of theories, there are $m$-brane solutions in $D=n+m+2$
dimensions with metric of the form
$$
 ds^2=H^{-\dd}(-dt^2+ dx_1^2+\dots+dx_m^2)+H^{\aa} (dr^2+ r^2 d\www _N^2),
\eqn\efopo$$ with 
$$   H=c+ {a ^{n-1}\over r ^{n-1}}
\eqn\erter$$ where $c,a,\aa,\dd$ are constants, $d\www _N^2$ is the metric on some
$n$-dimensional space $N$ ($N$ is an $n$-sphere $S^n$ for the most symmetric
solutions, and is   an Einstein space in typical cases). There may be a scalar
field with non-trivial $r$ dependence of the form
$$e^\phi = H^\gg
\eqn\dfsjhk$$ for some constant $\gg$. Near $r=0$,  which is a horizon if
$\dd>0$, the constant term in $H$ can be dropped and the metric becomes conformal
to a metric $d\ti s^2 $ on the product of $m+2$ dimensional AdS space and 
$N$, 
$$ds^2 = H^{A} d\ti s^2,\qq A= \aa-{2\over n-1}
\eqn\confof$$ and
$$ d\ti s^2 ={r^\bb  \over  a^\bb}dx_{\vert \vert} ^2  
 +{a^2 \over r^2} d r^2+a^2 d  \www _N^2
\eqn\abc$$ where  the longitudinal metric is $dx_{\vert \vert} ^2  =-dt^2+
dx_1^2+\dots+ dx_m^2$ and $\bb= 2-(n-1)(\aa+\dd)$. For some physical questions,
it is more natural to work with the \lq dual frame' metric
$d\ti s^2$ rather than $ds^2$ [\peet]. The change of variables $U=r^{\bb/2}$
brings the metric on the AdS space to the form
$$ds_{AdS}^2
 ={U^2  \over  a^\bb}dx_{\vert \vert} ^2  +
 {4a^2 \over \bb^2}{1 \over U^2} dU^2 
\eqn\adsm$$ with dilaton
$$e^\phi \propto U^{-2(n-1)\gg/\bb}
\eqn\abc$$ The constants in \adsm\ can be absorbed into rescalings of the
coordinates. For the space \efopo, as $r$ becomes large, $H$ approaches the constant
$c$ and the space approaches flat space $\R^D$ if $N$ is a round sphere $S^n$, or
the product of 
$m+1$ dimensional Minkowski space and a cone over $N$ otherwise.

This interpolating brane solution is the basis for arguments  that the theory in
the \lq near horizon geometry'  
\adsm\ is holographically dual to   an $m+1$ dimensional theory, which is the
world-volume theory of the brane. For the D3,M2 and M5 branes, $A=0$ and  the
scalar is either absent or constant, and the  dual theory is a superconformal
field theory. The solution is invariant under the AdS group $SO(m-1,2)$
 and the $SO(n+1)$ isometry group of $N=S^n$, and these, together with the 32
supersymmetries, form  a super-AdS group which is interpreted as the
superconformal group of the dual conformal field theory. For other branes, with
$N=S^n$ but 
$\gg\ne 0$, the near horizon limit is again
$AdS_{m+2}\times S^n$ in the dual frame, but the $r$-dependence of the dilaton
breaks the AdS group down to a Poincar\' e group, and the  dual theory is
non-conformal, with the  dependence of the string coupling on the radial
coordinate $r$ or $U$  corresponding to a scale dependence of the dual field
theory coupling constant.  The $SO(n+1)$ remains as an R-symmetry of the dual
theory. In the more general case in which $N$ is not a sphere, the dual theory is
the world-volume theory of a cone at a conical singularity, with $R$-symmetry
given by the isometry group of $N$. Note that this   formally extends to the
cases in which $N$ is non-compact or has a metric of non-Euclidean signature.

These solutions interpolating between $AdS\times N$ and flat space (or a cone)
have analogues for solutions interpolating between a de Sitter solution
$dS\times   N$ and flat space or a generalised cone. A similar statement may also
be true of the warped products  of section 2, but here attention will be
restricted to the  direct product solutions of sections 3 and 4, for which the
interpolating solutions and some generalisations have been given in
[\tim,\time,\HK]; these solutions   generically preserve some of the
supersymmetries.

 Consider then a brane-like space in $D=n+m+1$ dimensions with metric  
$$
 ds^2=H^{-\dd}( dx_1^2+\dots+dx_m^2)+H^{\aa} (-dt^2+dr^2+ r^2 d\www _{ N}^2),
\eqn\efopoe$$ where $H$ is a function of $r$ and $t$ in general, $c,\aa,\dd$ are
constants and 
$d\www _{ N}^2$ is the metric on some
$n-1$-dimensional space $ N$. While the solution \efopo\ is interpreted as
representing an  object extended in $m$ space and one time dimension that is
localised at
$r\sim 0$ and is a timelike surface parameterised by $t,x_1,\dots x_m$, the
solution
\efopoe\ represents a \lq Euclidean brane' which is an $m$-dimensional spacelike
surface  parameterised by $x_1,\dots x_m$.  It can be  localised in the
transverse space parameterised by $t,r $ and the coordinates of $ N$, with the
form of the \lq localisation' depending on the choice of $H$, which is a harmonic
function on the transverse space. As before there will in general  be a scalar
field of the form
$$e^\phi = H^\gg
\eqn\abc$$ for some constant $\gg$.

One class of solutions is that in which $H$ is independent of $t$ and given by
$$H=c+{q\over r^{n-2}}
\eqn\hhisfgh$$
 Such branes arise from time-like T-dualities; for example, starting with a
Dp-brane of type II string theory with $p=m$
 in a background with periodic time, then performing a T-duality in the time
direction gives an E-brane solution of the form \efopoe\  [\tim].
The T-dual of an AdS solution in the compact time direction is a solution of this
form with $c=0$, so that the constant term is absent in the harmonic function.

Another class of solutions has $H$   independent of $r$ and
given by a linear function of $t$,
$$H=mt+b
\eqn\abc$$ This gives a \lq cosmological' solution (which can be viewed as an
analogue of the domain-wall spacetimes which are of the form \efopo,\dfsjhk\ with
$H$ having a linear dependence on one of the spatial coordinates). The theories of
[\tim,\time] have many supersymmetric solutions of this form.

Next, there are solutions in which $H$ depends on the proper time $   \tt^2 = 
t^2-r^2 
$
$$   H=c+ {a^{n-1}\over \tt ^{n-1}}
\eqn\hhis$$ where for regularity in the region
$t^2>r^2$ we take $c\ge 0, a^n\ge 0$. Again it is useful to define a conformally
related dual frame metric 
$d\ti s ^2$  of the form \confof. The space with metric $d\ti s ^2$ and
coordinates restricted to the region
$\tt^2\ge 0$ is geodesically complete and non-singular [\tim].  As $\tt^2$
becomes large, $H$ tends to a constant and the space with metric $ds^2$ or
$d\ti s^2$ tends to flat space if
$ N$ is a round sphere, or to a product of $m+1$ dimensional Minkowski space
and a cone over  $ N$ otherwise.
 Near $\tt=0$, the constant term in
$H$ can be neglected so that
$H \sim (a/\tt)^{n-1}$. To study the behaviour near $\tt=0$, it is useful to
define  the Rindler-type coordinates $\tt,\rho$ by
$$ t= \tt \cosh \rho, \qq r =  \tt \sinh \rho 
\eqn\trtryru$$ so that
 the metric $d\ti s ^2$ becomes
$$
 d\ti s^2={\tt ^\bb  \over  a^\bb}dx_{\vert \vert} ^2  
 -{a^2 \over \tt^2} d \tau^2+a^2 d\ti \www _{\ti N}^2,
\eqn\abc$$ where $ d \www _{\ti N}^2$ is the metric on the $n$-dimensional space
${\ti N}$ with metric
$$ d  \www _{\ti N}^2= d \rho ^2 + \sinh ^2 \rho \, d \www _{ N}^2
\eqn\tinis$$ If $ N$ is a round $n-1$ sphere $S^{n-1}$, then
${\ti N}$ is the hyperbolic space $H^n$, the coset space $SO(n,1)/SO(n)$, with
\lq radius'  $1$. The longitudinal metric is now the Euclidean metric
$ dx_{\vert \vert} ^2 =dx_1^2+\dots +dx_m^2$. Then  
$$ds^2={\tt ^\bb  \over  a^\bb}dx_{\vert \vert} ^2  
 -{a^2 \over \tt^2} d \tau^2
\eqn\abc$$ is   the metric on $m+1$ dimensional de Sitter space, $dS_{m+1}$.
The change of variables $T=(\bb/2a^{1+\bb/2})\tt^{\bb/2}$ brings the metric on
the  dS space to the form $(4/\bb^2)ds_{dS}^2$ where
$$ds_{dS}^2
 ={T^2  \over  a^2}dx_{\vert \vert} ^2  
 -{a^2    \over T^2} dT^2 
\eqn\abc$$ with dilaton
$$e^\phi \propto T^{-2(n-1)\gg/\bb}
\eqn\abc$$

In the limit $c = 0$,
 the coordinates   $T ,  x_{\vert \vert} $ with  $T >0$ cover only   half of   de
Sitter space. There is a coordinate singularity at $T=0$,
 and the geometry can be continued through this to give the complete non-singular
de Sitter solution, with $T=0$ the cosmological horizon. The transverse space
has metric $ -dt^2+dr^2+ r^2 d\www _{ N}^2 $ and the region
$t^2>r^2$ consists of the interior of the past and future light cones of the
origin.
 The interior of the light-cone splits into two regions, the past light-cone
$t<r<0$ and the future light-cone
$0<r<t$, and  it is natural to define the proper time so that these are the two
regions
$T<0$ and $T>0$, and correspond to the two halves of the de Sitter
space [\tim]. For the  metric $d\ti s^2$ with
$t^2>r^2$, the region near
$t^2=r^2$ or $\tt=0$ is described by a non-singular $dS \times \ti N$ geometry, 
and
$\tt \to -\tt $ is an isometry, so that one can argue as in [\dilres] that the 
space can be continued through the coordinate singularity at $\tt =0$. Then   
the region in which $\tt$ is  real or
$t^2>r^2$ of the   brane solution is also a complete non-singular solution. The
behaviour of the dilaton at $\tt=0$ will depend on the coefficients
$\bb,\gg$, but in many   cases  (including the
E4-brane, which has constant dilaton), it can be continued smoothly through
$T=0$.

There are also 
solutions \efopoe\
with $$   c'+ {b^{n-1}\over \ss ^{n-1}}
\eqn\hhiss$$ where  the  proper distance is $   \ss^2 = r^2 - t^2
$ that are regular for $t^2<r^2$  with $c',b^{n-1}$ real and non-negative. 
If $n-1$ is a multiple of 4, this a continuation of the
 solution \efopoe,\hhis\ to   the region $t^2<r^2$, but for other $n$ it should be
regarded as a distinct solution. The analysis is similar to that of the region
$\tt^2>0$. In the region
$r^2> t^2$, we use   \efopoe\ with \hhiss\
 and define the  coordinates $\ss,\xi$ by
$$ r= \ss \cosh \xi, \qq t =  \ss \sinh \xi 
\eqn\dssad$$ so that the metric   becomes  
$$
 ds^2={\ss ^\bb\over a^\bb}  dx_{\vert \vert} ^2  
 +{b^2 \over \ss^2} d \sigma^2+b^2 d\hat  \www _{\bar N}^2
\eqn\ebrsp$$ where
$$ d\bar\www _{\bar N}^2= -d \xi ^2 + \cosh ^2 \xi  d \www _{  N}^2
\eqn\tiniss$$ 
If $  N$ is a round
$n-1$ sphere, then this
 is the $n$-dimensional de Sitter metric of \lq radius' $1$, while more
generally it is a de Sitter-like cosmology with spatial section $N$ instead of
$S^{n-1}$.  The metric  $$ds^2={\ss ^\bb\over b^\bb}  dx_{\vert \vert} ^2  
 +{b^2 \over \ss^2} d \sigma^2
\eqn\abc$$ is   the   metric on the hyperbolic space $H^{m+1}$:  the change of
variables $X=(\bb/2b^{1+\bb/2})\ss^{\bb/2}$ brings the metric     to the
standard metric on $H^{m+1}$  
$ds^2=(4/\bb^2)ds_{H}^2$ where
$$ds_{H}^2
 ={X^2  \over  b^2}dx_{\vert \vert} ^2  
 +{b^2    \over X^2} dX^2 
\eqn\abc$$ with dilaton
$$e^\phi \propto X^{-2(n-1)\gg/\bb}
\eqn\abc$$
The boundary of $H^{m+1}$  is the sphere $S^{m}$ given by the hyperplane
$X=0$ (plus a point at infinity). In the brane solution \efopoe, it is at an
infinite geodesic distance, with respect to $d\ti s^2$, from any interior
point  to the boundary
$X=0$ (corresponding to
$\ss=0$ if
$\bb>0$ and
$\ss=\infty$ if $\bb<0$), and again the solution is complete, although the
dilaton can blow up at the boundary if $\gg/\bb>0$.

To summarise, the metric \efopoe\ has two regions, $\tt^2>0$ and $\tt^2 <0$,
and each region can be a complete space.
This complete space typically interpolates between a \lq near-horizon
geometry' $X$ and an asymptotic region where $H \sim 1$, which is 
Minkowski
space in $n+m+1$ dimensions $\R^{n+m,1}$  when $N$ is
a sphere $S^{n-1}$, and more generally is the product of
$\R^{m,1}$ and a cone over $N$. 
 If $H$ is of the form \hhis, then the
region $\tt^2\ge 0$ is a complete space whose near-horizon
geometry is $X=dS_{m+1}\times H^n$ when $N$ is a sphere $S^{n-1}$, and more
generally is the product of $dS_{m+1}$ and the space $\ti N$ with metric \tinis.
 If $H$ is of the form \hhiss, then the
region $\tt^2<0$ is a complete space whose near-horizon
geometry is $X=H^{m+1}\times dS_n$ when $N$ is a sphere $S^{n-1}$, and more
generally is the product of $H^{m+1}$ and the space $\hat N$ with metric
\tiniss.

\chapter {Holography}

In [\mal],   $N$ parallel D$3$-branes separated by distances of order $\rr$  were
considered and   the zero-slope limit
$\aa '\to 0$  was taken keeping $r=\rr / \aa '$ fixed, so that the energy of
stretched strings remained finite. This decoupled the bulk and string degrees of
freedom  leaving a theory on the brane which is $U(N)$ ${\cal N}=4$ \sym\ with
Higgs expectation values, which are of order $r$, corresponding to the brane
separations.   The  D$3$-brane supergravity solution is of the form
\efopo,\erter\ and
 has charge $q=a^2 \propto Ng_s/{\aa '}^2$ where $g_s=g_{YM}^2$ is the string
coupling constant and $g_{YM}$
 is   the \sym\ coupling constant. Then as $\aa '\to 0$, $q$ becomes large and the
background becomes $AdS_5\times S^5$. The IIB string theory in the $AdS_5\times S^5$
background is a good description if the curvature $R \sim 1/a^2$ is not too
large, while if $a^2$ is large, the \sym\ description is reliable.  In the 't
Hooft limit in which $N$ becomes large while $g_{YM}^2N$ is kept fixed, $g_s \sim
1/N$, so that as $N\to \infty$, we get the free string limit $g_s\to 0$, while
string loop corrections correspond to $1/N$ corrections in the \sym\ theory.
The energy-scale of the \sym\ theory 
is associated with the radial coordinate $r$ of the $AdS$ space, and going to the
boundary $r \to \infty$ in the AdS space coresponds to taking the  
ultra-violet limit of the \sym\ theory. The \sym\ theory with UV cut-off $\lll$
can in some ways be thought of as being located at  a surface $r=r_0$, with the
constant $r_0$ tending to infinity in the limit $\lll \to \infty$.

Similar arguments apply here, with the two   E$4$-brane solutions, 
given by \efopoe\ and \hhis\ or \hhiss\  with $n=3$, $m=5$, $N=S^3$ and
constant $\phi$, with spacelike
or timelike interpolations corresponding to whether the separation between the 
E$4$-branes that is kept fixed is spacelike or timelike. Recall that the scalars
of the \sym\ theory are in a vector representation of the $SO(5,1)$ R-symmetry,
where those in the {\bf 5} of
$SO(5)\subset SO(5,1)$ have kinetic terms of the right sign and correspond  to
brane separations in the 5 spacelike 
 transverse dimensions, while the remaining ($U(N)$-valued) scalar  has
 kinetic terms of the \lq wrong' sign
and corresponds to timelike separations of the E-branes.

Consider first the case of   $N$ parallel E$4$-branes of the $IIB^*$ string
theory separated by distances of order $\rr$  in one of the 5 spacelike
transverse dimensions. We take the zero-slope limit
$\aa '\to 0$ keeping $\ss=\rr / \aa '$ fixed, so that the energy of stretched
strings remains  finite. This gives a decoupled theory on the brane consisting of
the $U(N)$ ${\cal N}=4$ Euclidean \sym, with Higgs expectation values of order
$\ss$ for the scalars   corresponding to the spacelike separations.  The
corresponding supergravity background is the E$4$-brane with spacelike
interpolation and $H$ given by \hhiss, arising from the outside of the
light-cone with $\ss$ real and positive.  We again have  $q=a^2 \propto
Ng_s/{\aa '}^2$ and $g_s=g_{YM}^2$, so that for large $N$, the system can be 
described by the $IIB^*$ string theory in $dS_5\times H^5$ if $a^2$ is large
and by the large $N$ Euclidean \sym\ theory when
$a^2$ is small. In the 't Hooft limit,  string loop corrections again correspond
to $1/N$ corrections in the \sym\ theory. The Euclidean gauge theory
can be thought of as located at the boundary $S^4$ of $H^5$.

For   $N$ E$4$-branes of the $IIB^*$ string theory separated by distances of
order $T$  in the timelike transverse dimension,  we take the zero-slope limit
$\aa '\to 0$ keeping $\tt=T/ \aa '$ fixed. This gives a decoupled theory on the
brane consisting of the $U(N)$ ${\cal N}=4$ Euclidean \sym, with Higgs
expectation values of order $\tt$ for the scalars   corresponding to the timelike
separations.  The corresponding supergravity background is the E$4$-brane with
timelike interpolation with $H$ given by \hhis, arising from the inside of the
light-cone with $\tt$ real. Again  for large $N$, the system can be  described
by the $IIB^*$ string theory in $dS_5\times H^5$ if $a^2$ is large and by the
large $N$ Euclidean \sym\ theory when
$a^2$ is small. In this case, the Euclidean gauge theory can be thought of as 
being located at the past (or future) Cauchy surface $\tt= -\infty$ ($\tt= 
\infty$). 
It was pointed out in [\strom]
that  the past and future boundaries of de Sitter
space can be identified by identifying
  points connected by null geodesics, and that this is natural in the context of de
Sitter holography,  so that the holographic dual is a field theory on one
4-dimensional boundary, rather than two. 
The identified pair of boundaries will be referred to as \lq the boundary'.

The four dimensional Euclidean \sym\  theory has scalars taking values in
a Lorentizan space $\R^{5,1}$. The vacua split into two branches, depending on
whether the expectation value of the scalar fields $v^i=<\phi^i>$ is spacelike
 or
timelike. The branch with $v^2>0$
corresponds to E4-branes that are spacelike separated in 10-dimensions and 
arises as a  holographic theory on the boundary of $H^5$, with the scale of the CFT
related to  a spatial radial coordinate on $H^5$, while the 
branch with $v^2<0$
corresponds to E4-branes that are timelike separated   and 
arises as a  holographic theory on the boundary of $dS_5$, with the scale of
the CFT related to  a timelike   coordinate on $dS_5$.
If the $H^5$ is compactified by identifying under a discrete isometry group,
the CFT is identified under the corresponding discrete subgroup of the R-symmetry
group, and only the branch with $v^2<0$ remains, as the dual theory on the boundary
of de Sitter space.

Note that in the above, the dimension of the de Sitter space and of the hyperbolic
space are the same. Consider cases in which  this generalises to $d$ dimensional
Euclidean CFTs with conformal group 
$SO(d,1)$ and R-symmetry group $SO(n,1)$, with $n+1$ scalars
taking values in the Lorentzian space $\R^{n,1}$.
For spacelike scalar expectation values, the   dual theory would arise from
a theory on $H^d\times dS_n$ as a theory on the boundary of $H^d$, while for
timelike  scalar expectation values, the dual theory would be in
$dS_d\times H^n$, with the CFT arising on the boundary of the de Sitter space.
The theories on the boundaries of $dS_n$ or $H^n$ in these two cases
would then be two branches of an $n$ dimensional conformal field theory with
$SO(d,1)$ R-symmetry, giving an interesting chain of dualities.
Explicit supersymmetric examples of this kind were given in [\HKt].

As a further example, consider as in [\mal] the   type IIB string with a set of 
$N$ parallel D5-branes wrapping a torus $T^4$, with
$M$  parallel D1-branes lying in the non-compact direction of the
D5-branes.
The near-horizon geometry is 
$AdS_3\times S^3$ and the bulk theory has 16 supersymmetries.
The 2-dimensional dual CFT has (4,4) supersymmetry and $SO(4)$ R-symmetry.
This consists of a $D=2$ \sym, obtained
from reducing
$N=2,D=6$
\sym\ on
$T^4$, and   a supersymmetric sigma-model whose target space is the
$M$-instanton moduli space   in $SU(N)$ Yang-Mills [\mal].
Now, if time is also compact, then T-dualising the D1-D5 system in the  time
direction gives an E2-E6 brane system of the type $IIA^*$ theory
wrapped on $T^4$, with
the E2-branes lying in the non-compact directions of the E6-branes.
The near horizon geometry is now $dS_3\times H^3$
and the corresponding 2-dimensional Euclidean conformal
field theory has R-symmetry $SO(3,1)$ and (4,4) supersymmetry. (For a discussion of
$(p,q)$ supersymmetry in 2 Euclidean dimensions, see [\abz].) The 2-dimensional
Euclidean \sym\ theory with $SO(3,1)$ R-symmetry is that obtained by reducing
$N=2,D=6$
\sym\ on a Lorentzian torus
$T^{3,1}$, while the sigma-model again has the same instanton moduli space as its
target space, but the world-sheet is now Euclidean.

As for  the AdS case, it seems that the holographic duality between 
a bulk theory
in de Sitter space and a Euclidean conformal field theory that formally 
arises in   supersymmetric cases applies more generally. Indeed, as was
seen in section 8,   de Sitter solutions often have associated Euclidean brane
solutions interpolating between them and flat space or a conical spacetime, and
these can  form the basis for an argument in the style of [\mal].
 Further evidence
for such a de Sitter holography  has   been discussed in
[\strom,\klemm].  

It is natural to ask whether such a holographic duality can extend  
to de Sitter solutions  
of conventional supergravities of the type discussed in section 2.
It seems plausible that for these solutions too there should be 
spacelike brane  solutions interpolating between flat space and the de Sitter space 
solution which could form the basis of a Maldacena-style argument. However, in this
case, all supersymmetries are broken and the usual issues arise as to how far one
can trust such arguments in the absence of supersymmetry.

\refout

\bye